\documentclass[%
 reprint,
 amsmath,amssymb,
 aps,
]{revtex4-2}

\usepackage[dvipsnames]{xcolor}
\usepackage{graphicx}
\usepackage{dcolumn}
\usepackage{bm}
\usepackage{tikz}
\usepackage{pgfplots}
\usepackage{subfigure}
\usepackage{multirow}
\usepackage{diagbox}
\usepackage{makecell} 
\usepackage{booktabs} 
\usepackage{float}
\usepackage{amsthm}
\usepackage{physics}
\usepackage{tabularray}
\UseTblrLibrary{booktabs}
\UseTblrLibrary{diagbox}

\definecolor{babyblue}{RGB}{131,195,221}
\definecolor{lightblue}{RGB}{48,155,200}
\definecolor{justblue}{RGB}{1,102,169}
\definecolor{deepblue}{RGB}{2,52,107}

\definecolor{babyred}{RGB}{254,134,110}
\definecolor{lightred}{RGB}{252,67,61}
\definecolor{justred}{RGB}{201,24,40}
\definecolor{deepred}{RGB}{111,3,25}

\usepackage[colorlinks,
linkcolor=red,
anchorcolor=blue,
citecolor=green
]{hyperref}

\DeclareMathOperator{\Rank}{Rank}

\begin{document}

\title{Learning Symmetric Hamiltonian}

\author{Jing Zhou}

\affiliation{Institute of Physics, Beijing National Laboratory for Condensed
  Matter Physics,\\Chinese Academy of Sciences, Beijing 100190, China}
	
\affiliation{School of Physical Sciences, University of Chinese Academy of
  Sciences, Beijing 100049, China}

\author{Le-Yi Lu}

\affiliation{Institute of Physics, Beijing National Laboratory for Condensed
  Matter Physics,\\Chinese Academy of Sciences, Beijing 100190, China}

\affiliation{School of Physical Sciences, University of Chinese Academy of
  Sciences, Beijing 100049, China}

\author{D. L. Zhou} \email[]{zhoudl72@iphy.ac.cn}
	
\affiliation{Institute of Physics, Beijing National Laboratory for Condensed
  Matter Physics,\\Chinese Academy of Sciences, Beijing 100190, China}
	
\affiliation{School of Physical Sciences, University of Chinese Academy of
  Sciences, Beijing 100049, China}

\date{\today}

\begin{abstract}
Hamiltonian learning is a process of recovering system Hamiltonian from measurements, which is a fundamental problem in quantum information processing. In this study, we investigate the problem of learning a symmetric Hamiltonian from its eigenstate. By the application of group representation theory, we have derived a method to determine the number of linearly independent equations about the Hamiltonian unknowns obtained from an eigenstate. Without accidental symmetry, the Hamiltonian recoverability is determined by the degeneracy of the associated irreducible representation of the Hamiltonian symmetry group. To illustrate our approach, we examine the XXX Hamiltonian and the XXZ Hamiltonian. We first determine the Hamiltonian symmetry group, then work out the decomposition of irreducible representation, which serves as foundation for analyzing the uniqueness of recovered Hamiltonian. We also take the XXX Hamiltonian as an accidental symmetric case of the XXZ Hamiltonian to investigate the Hamiltonian recovery with an accidental symmetry. Our numerical findings consistently align with our theoretical analysis.
\end{abstract}

\maketitle

\section{introduction}

Hamiltonian is the characteristic function of a quantum system, which contains all the information on the system in principle. Hamiltonian Learning is an information task to recovering the Hamiltonian of a quantum system, typically from observations of its steady state or dynamics. Hamiltoninan learning, as an essential tool for verification and benchmarking of a quantum system, plays a pivotal role in modern quantum technology. In recent years, there has been a rapid development in quantum simulators and related quantum computing devices, including controllable trapped ion~\cite{2011Quantum, PhysRevLett.121.180501, PhysRevLett.117.060504, PhysRevLett.74.4091} and superconducting quantum circuits~\cite{PhysRevLett.111.080502,Barends2014}. The exponential growth of noises with the size of quantum hardware has made Hamiltoninan learning of these systems full of challenge~\cite{PRXQuantum.3.030324,PhysRevA.101.062305,wang2017experimental, motta2020determining, wiebe2014hamiltonian,PhysRevA.91.042129}. In the field of condensed matter physics, Hamiltoninan learning aids in understanding the behaviors of quantum material and may be utilized to explore new material systems~\cite{PhysRevB.99.235109,PhysRevLett.130.200403, anshu2021sample, shabani2011estimation, PhysRevA.79.020305,PhysRevLett.127.200503,9317911}.

One of the challenges in Hamiltoninan learning lies in determining when we can uniquely recover a Hamiltonian from its steady state. This question was initially raised and partially addressed by Qi and Ranard, and based on the energy variance is zero for any eigen energy state they show that most of the generic local Hamiltonian can be recovered from measurements of a single eigenstate~\cite{Qi2019determininglocal}. Along this research direction, various methods for recovering the Hamiltonian from a steady state have been proposed~\cite{PhysRevX.8.031029,PhysRevLett.122.020504,Hou_2020}. In particular, based on the steady state density operator commutates with the Hamiltonian, Bairey, Arad, and Linden develop an effective algorithm to recover Hamiltonian from local observables~\cite{PhysRevLett.122.020504}.

To further clarify how many independent equations for unknown parameters of Hamiltonian can be obtained from a single steady state, we develop an equivalent approach based on the energy eigen equation in Ref.~\cite{PhysRevA.105.012615}, where we can give the analytical results on the number of independent equations from a single eigen state in spin chain models, and correctly predict the local Hamiltonian can be uniquely recovered up to a constant only when the spin chain exceeds a critical length. Subsequently, we extended this investigation to degenerate steady states using the orthogonal space equation method~\cite{ZHOU2024129279}.

So far, we focused on the recoverability of a generic local Hamiltonian from its steady state, while many Hamiltonians arousing interests exhibit symmetries~\cite{doi:10.1073/pnas.93.25.14256,PhysRevLett.129.160503}, such as the Heisenberg spin chain models~\cite{WOS:000319298800007}. Symmetric Hamiltonians inherently entail fewer unknowns compared to generic ones. However, it doesn't necessarily make them easier to recover. According to Noether's theorem, there is a direct relationship between the Hamiltonian symmetry and the conserved quantities in the associated physical system. It directly reduces the degrees of freedom of associated eigenstates~\cite{kosmann2011noether}. This is easy to understand, because every eigenstate, from which we try to recover the Hamiltonian, must belong to some smaller space in which the equations can provide non-trivial information about Hamiltonian.

In this paper, we will extend the research realm from a generic Hamiltonian learning to a symmetric Hamilton learning. First, we will define the symmetry group in the context of Hamiltonian learning. By the aid of the powerful group representation theory, we extend our eigen energy equation approach~\cite{PhysRevA.105.012615} to the symmetric case, where the irreducible representation subspaces are essential to obtain independent equations. We also discuss the appearance of accidental symmetry and how it affects symmetric Hamiltonian learning. To verify our theoretical derivation, we perform simulations on the XXX and XXZ Hamiltonians. Given only the Hamiltonian's symmetry group and its irreducible representations, we are able to successfully predict the number of linearly independent equations obtained from an eigenstate, which is directly related with the Hamiltonian recovery.

\section{Symmetric Hamiltonian learning problem}
\label{sec:symm-hamilt-learn}

Let us formulate our problem as follows. Our quantum system is described by a Hamiltonian in a $d$-dimensional Hilbert space $\mathcal{H}$:
\begin{equation}
\label{Hamiltonian}
    H(\{a_n\}) =\sum_{n=1}^N a_{n}h_{n},
\end{equation}
where $a_n$ is a real parameter,  $h_n$ is a traceless Hermitian operator, any two different operators $h_n$ and $h_m$  are linearly independent, and $N$ is the number of linearly independent terms in the Hamiltonian $H$. In our problem, all operators $\{h_n\}$ are known, but all parameters $\{a_n\}$ are unknown, and to be learned from the knowledge coming from measurements.

To be concrete, we assume that we know a single eigen state $|\psi\rangle$ of the Hamiltonian $H(\{a_n^{\ast}\})$, i.e., $|\psi\rangle$ satisfies the eigen equation
\begin{equation}
  \label{eq:4}
  H(\{a_n^{\ast}\}) |\psi\rangle = E |\psi\rangle,
\end{equation}
where the eigen energy $E$ is unknown. The task is to recover all the unknown parameters $\{a_n^{\ast}\}$ up to a real constant multiplier. This problem is first advocated by Qi and Ranard in Ref.~\cite{Qi2019determininglocal}.

Note that there are $d^2-1$ linearly independent traceless Hermitian operators in $\mathcal{H}$. If the number $N=d^2-1$, then all kinds of Hamiltonians are possible, there are infinitely many Hamiltonians with $|\psi\rangle$ being one energy eigen state, and we are impossible to recover the Hamiltonian from a single energy eigen state. Fortunately,  in almost all practical problems, $N\ll d^2-1$, especially for a quantum many-body system. For example, when we consider a spin chain model, which consists $L$ $\frac{1}{2}$-spins along a line, only the nearest neighbor interactions or the next nearest interactions are needed to be introduced.

It is worthy to emphasize that in the above Hamiltonian learning problem there always exists a Hamiltonian $H(\{a_n^{\ast}\})$ with $|\psi\rangle$ being its eigen state, which implies that $H(\{c a_n^{\ast}\})$ with $c$ being any nonzero real parameter will also satisfy the requirement . The fundamental condition for such a recovery is that there does not exist any other Hamiltonian with $|\psi\rangle$ being its eigen state. The basic problem is: When this fundamental condition is satisfied and then how to recover the Hamiltonian correctly?

So far there are many elegant and powerful methods to recover a generic Hamiltonian from a single energy eigen state~\cite{Qi2019determininglocal,PhysRevX.8.031029,PhysRevLett.122.020504,Hou_2020,PhysRevA.105.012615,ZHOU2024129279}. However, in real systems, the Hamiltonians often have some symmetries.  A natural question arises: when the Hamiltonian is symmetric, what does the symmetry affect the problem?

First let us clarify the meaning of the symmetry of the Hamiltonian $H$ in Eq.~\eqref{Hamiltonian} with unknown parameters. The symmetry group for the Hamiltonian $H$ is defined by
\begin{equation}
  \label{eq:5}
  G_H = \{S: S H S^{\dagger} = H\},
\end{equation}
where $S$ is a unitary transformation or an antiunitary transformation. In the traditional quantum mechanics, the symmetry group is defined for a given Hamiltonian. In the Hamiltonian learning problem, however, it is more convenient to define the symmetry group for all the Hamiltonians specified by Eq.~\eqref{Hamiltonian}. Then all the parameters $\{a_n\}$ can be regarded as independent, which implies that Eq.~\eqref{eq:5} can be simplied as
\begin{equation}
  \label{eq:6}
  G_H = \{S: S h_n S^{\dagger} = h_n, \forall n\in\{1,2,\cdots,N\} \}.
\end{equation}

Note that the symmetry group defined by Eq.~\eqref{eq:6} contains all the common symmetry transformations for every Hamiltonian term in Eq.~\eqref{Hamiltonian}. In general, a Hamiltonian for a specific $\{a_n\}$ may have a higher symmetry accidentally, which contains more symmetry transformations. If a specific Hamiltonian with a higher symmetry, we call the Hamiltonian with accidental symmetry. Otherwise, we call the Hamiltonian without accidental symmetry. 

Now we are ready to define our problem: Our task is to recover $\{a_n^{\ast}\}$ up to a constant from a single eigen state $|\psi\rangle$ of $H(\{a_n^{\ast}\})$. Our main contribution is introducing the symmetry group in Eq.~\eqref{eq:6} for the unknown Hamiltonian family in Eq.~\eqref{Hamiltonian}, and examine the effect of the symmetry group on our Hamiltonian learning task.

\section{Principle of symmetric Hamiltonian learning}
\label{sec:impl-symm}

The basic strategy to recover Hamiltonian we develop in Refs.~\cite{PhysRevA.105.012615,ZHOU2024129279} is to solve directly the energy eigen equation
\begin{equation}
  \label{eq:8}
  H(\{a_n\}) |\psi\rangle = E |\psi\rangle,
\end{equation}
where $\{a_n\}$ and $E$ are unknown real numbers.

The power of symmetry lies in the fact that we can choose a suitable basis in the Hilbert space $\mathcal{H}$ according to the symmetry group defined in Eq.~\eqref{eq:6}. According to the reprentation theory of the symmetry group, the basis can be taken as $\{|\phi^p_{i m}\rangle\}$, where $p$ represents the irreducible representation $p$ of the symmetry group $G_H$, $i$ is the index that distinguishes the degeneracy of the irreducible representation, $m$ is the component index in the same representation. The complete relation of the base vectors $\{|\phi^p_{im}\rangle\}$ for the Hilbert space $\mathcal{H}$:
\begin{equation}
  \label{eq:9}
  \sum_{p=1}^{n_r} \nu_p d_p = d,
\end{equation}
where $n_r$ is the number of irreducible representations of $G_H$, $\nu_p$ is the degeneracy of the irreducible representation $p$, and $d_p$ is the dimension of the irreducible representation.

\subsection{The case of $H(\{a_n^{\ast}\})$ without accidental degeneracy}
\label{sec:case-ha_ncst-with}

In the case that $H(\{a_n^{\ast}\})$ does not have accidental degeneracy, the eigen state $|\psi\rangle$ must belong to some irreducible representation $p$, i.e.,
\begin{equation}
  \label{eq:10}
  |\psi\rangle = \sum_{im} c^p_{im} |\phi^p_{im}\rangle.
\end{equation}

From the definition of the symmetry group $G_H$ in Eq.~\eqref{eq:6}, we know that every $h_n$ is a scalar for $G_H$. According to the Wigner-Eckart theorem~\cite{hassani2013mathematical,cornwell1997group,zee2016group,XuGroupEn}, we obtain
\begin{equation}
  \label{eq:11}
  \langle \phi^q_{j l} | h_n |\phi^p_{i m}\rangle  = \delta_{pq} \delta_{ml} \langle \phi^p_j||h_n||\phi^p_i\rangle,
\end{equation}
where the matrix $\langle \phi^p_j||h_n||\phi^p_i\rangle$ is called the reduced matrix of $h_n$, which is independent of the indexes $l,m$. Eq.~\eqref{eq:11} implies that $H(\{a_n^{\ast}\})$ is block diagonalized in $\{|\phi^p_{im}\rangle, i\in\{1,2,\cdots,\nu_p\}\}$ for different $p$ and $m$, and it has the same matrix block for different $m$ under the same $p$. Thus the eigen states of energy $E$ of $H(\{a_n^{\ast}\})$ are $\nu_p$-folds:
\begin{equation}
  \label{eq:33}
  |\phi^p_m\rangle = \sum_i c^p_i |\phi^p_{im}\rangle,
\end{equation}
where the coefficient $c_i^p$ does not depend on $m$. Then the eigen state
\begin{equation}
  \label{eq:34}
  |\psi\rangle = \sum_m \lambda_m |\phi^p_m\rangle = \sum_{im} \lambda_m c^p_i |\phi^p_{im}\rangle,
\end{equation}
where $\lambda_m$ are any complex coefficient. Comparing Eq.~\eqref{eq:10} and Eq.~\eqref{eq:34}, we derive
\begin{equation}
  \label{eq:35}
  c^p_{im} = \lambda_m c^p_i.
\end{equation}

Inserting Eq.~\eqref{eq:10}, Eq.~\eqref{eq:11} and Eq.~\eqref{eq:35} into Eq.~\eqref{eq:8}, we obtain
\begin{equation}
  \label{eq:12}
  \sum_{n=1}^N a_n D_{n j}^p - E c_{j}^p = 0,
\end{equation}
where $j\in\{1,2,\cdots,\nu_p\}$, and
\begin{equation}
  \label{eq:13}
  D_{n j}^p = \sum_{i=1}^{\nu_p}  \langle \phi^p_j||h_n||\phi^p_i\rangle c^p_i.
\end{equation}
Because the coefficients $D^p_{nj}$ and $c^p_{j}$ are complex in general, Eq.~\eqref{eq:12} is equivalent to two real equations:
\begin{align}
  \label{eq:15}
  \sum_{n=1}^N a_n \Re(D_{n j}^p) - E \Re(c_{j }^p) & = 0, \\
  \label{eq:16}
  \sum_{n=1}^N a_n \Im(D_{n j}^p) - E \Im(c_{j }^p) & = 0.
\end{align}
In the matrix form, Eq.~\eqref{eq:15} and Eq.~\eqref{eq:16} can be rewritten as
\begin{equation}
  \label{eq:17}
  Q^p_{j} \vec{x} = 0,
\end{equation}
where
\begin{equation}
  \label{eq:18}
  Q^p_{j} =
  \begin{pmatrix}
    \Re(D_{1 j}^p) & \Re(D_{2 j}^p) & \cdots & \Re(D_{N j}^p) & - \Re(c_{j }^p) \\
    \Im(D_{1 j}^p) & \Im(D_{2 j}^p) & \cdots & \Im(D_{N j}^p) & -\Im(c_{j }^p)
  \end{pmatrix}
\end{equation}
and $\vec{x}^T=(a_1, a_2, \cdots, a_N, E)$ with index $T$ denoting vector transposition. Note that Eq.~\eqref{eq:17} are valid for all possible $j$. Thus we can combining $Q^p_{j}$ into a single $Q^p$, and obtain
\begin{equation}
  \label{eq:19}
  Q^p \vec{x} = 0,
\end{equation}
where $Q^p$ is $(2\nu_p)\times(N+1)$ matrix.

It is worthy to note that in numerical calculations, we rewrite Eq.~\eqref{eq:19} as
\begin{equation}
  \label{eq:1}
  \lambda_m Q^p \vec{x} = 0, \forall \lambda_m \neq 0.
\end{equation}
Then we only to replace $c_i^p$ with $c_{im}^p=\langle\phi_{im}^p|\psi\rangle$ ranging from Eq.~\eqref{eq:12} to Eq.~\eqref{eq:18}.

Note that if $\vec{x}^{\ast}$ is a solution of Eq.~\eqref{eq:19}, then $c \vec{x}^{\ast}$ is also a solution for any nonzero real parameter $c$. Then the rank
\begin{equation}
  \label{eq:20}
  \Rank Q^p \le N.
\end{equation}
The condition of recovering the Hamiltonian up to a constant is
\begin{equation}
  \label{eq:21}
  \Rank Q^p = N.
\end{equation}
In addition, the rank must less and equal to the number of equations, i.e., $\Rank Q^p \le 2 \nu_p $. Then a necessary condition of recovering a Hamiltonian up to a constant is
\begin{equation}
  \label{eq:22}
  2 \nu_p  \ge N.
\end{equation}
Without accidental degeneracy, the above condition~\eqref{eq:22} is also sufficient for the recovery since we have any additional reason to reduce the number of independent equations.

\subsection{The case of $H(\{a_n^{\ast}\})$ with accidental degeneracy}
\label{sec:case-h-with}

In the case that $H(\{a_n^{\ast}\})$ with accidental degeneracy,  $H(\{a_n^{\ast}\})$ has more symmetry transformations than those in $G_H$, and its eigen state $|\psi\rangle$ can belong to different irreducible representations of $G_H$, i.e.,
\begin{equation}
  \label{eq:23}
  |\psi\rangle = \sum_{p\in \Lambda(\psi)} \sum_{im} c^p_{im} |\phi^p_{im}\rangle,
\end{equation}
where the set $\Lambda(\psi)=\{p: \exists i,m, \qq{s. t.} \langle\phi^p_{im}|\psi\rangle\neq 0\}$. It is easy to find that Eqs.~\eqref{eq:11}-\eqref{eq:19} are still valid. Then we combine the set of matrix $\{Q^p, p\in \Lambda(\psi)\}$ into a single matrix $Q^{\psi}$, whose size is $(\sum_{p\in \Lambda(\psi)} 2 \nu_p ) \times (N+1)$, and the constraint equations become
\begin{equation}
  \label{eq:27}
  Q^{\psi} \vec{x} = 0.
\end{equation}

Similarly, the rank
\begin{equation}
  \label{eq:24}
  \Rank Q^{\psi} \le N.
\end{equation}
The condition of recovering the Hamiltonian up to a constant is
\begin{equation}
  \label{eq:25}
  \Rank Q^{\psi} = N.
\end{equation}
The rank $\Rank Q^{\psi}\le \sum_{p\in \Lambda(\psi)} 2 \nu_p $, and a necessary condition of recovering the Hamiltonian up to a constant becomes
\begin{equation}
  \label{eq:26}
  \sum_{p\in \Lambda(\psi)} 2 \nu_p  \ge N.
\end{equation}
In general, the condition~\eqref{eq:26} may not be sufficient, which is due to the lack of the precise knowledge of the symmetry.

It is worthy to emphasize that if we randomly generate the Hamiltonian $H(\{a_n^{\ast}\})$ in Eq.~\eqref{Hamiltonian}, then the probability measure of generating Hamiltonian $H(\{a_n^{\ast}\})$ with accidental degeneracy is zero.

\subsection{Numerical algorithm for symmetry Hamiltonian learning}
\label{sec:numer-algor-symm}

In both cases of $H(\{a_n^{\ast}\})$ with or without accidental degeneracy, Eq.~\eqref{eq:19} and Eq.~\eqref{eq:27} can be written in a unified form
\begin{equation}
  \label{eq:28}
  Q \vec{x} = 0.
\end{equation}
The condition of recovering the Hamiltonian $H(\{a_n^{\ast}\})$ up to a constant is
\begin{equation}
  \label{eq:29}
  \Rank Q = N.
\end{equation}
In addition, we know that $Q^{\dagger}Q$ is a semipositive real symmetric $(N+1)\times(N+1)$ matrix, which can be diagonalized via a rotation matrix $R$ of $SO(N+1)$:
\begin{equation}
  \label{eq:30}
  R Q^{\dagger} Q R^T = D,
\end{equation}
where $D$ is a diagonalized matrix
\begin{equation}
  \label{eq:31}
  D =
  \begin{pmatrix}
    D_{11} & 0 & \cdots & 0\\
    0 & D_{22} & \cdots & 0 \\
    \vdots & \vdots & \vdots & \vdots \\
    0 & 0 & \cdots & D_{N+1,N+1}
  \end{pmatrix},
\end{equation}
where $D_{n,n}\ge D_{n+1,n+1}$ and $D_{N+1,N+1}=0$. Then the Hamiltonian recovering condition becomes $D_{N,N}>0$. In other words,  the Hamiltonian can be recovered up to a real multiplier if and only if $\vec{x}$ is the unique eigenvector of $Q^\dagger Q$ with eigenvalue $0$.

In addition, solving Eq.~\eqref{eq:28} can be transformed into a constraint optimization problem:
\begin{equation}
  \label{eq:32}
  \min_{\vec{x}} \vec{x}^T Q^{\dagger} Q \vec{x}
\end{equation}
with $\vec{x}^T=(a_1,\cdots,a_N,E)$ and $\sum_{n=1}^N a_n^2=1$. This optimization can be realized by the traditional numerical optimization algorithm.

\section{Examples demonstrating symmetric Hamiltonian learning}

Now we apply the general approach of symmetric Hamiltonian learning  to study two $1$-dimensional spin chain models, the XXX model and the XXZ model~\cite{Sachdev_2011,Franchini_17}. On one hand, it gives the explicit steps to implement symmetric Hamiltonian learning. On the other hand, it shows why symmetry is important in Hamiltonian learning problem.

\subsection{The XXX Spin Chain}\label{sec:xxx-spin-chain}

Considering an 1-dimensional spin chain of $L$ sites with a spin $\frac{1}{2}$ particle at each site. The XXX Hamiltonian model can be written as
\begin{equation}
H_{XXX} = \sum_{l=1}^{L-1}J_{l}(\sigma_{l}^{x}\sigma_{l+1}^{x}+\sigma_{l}^{y}\sigma_{l+1}^{y}+\sigma_{l}^{z}\sigma_{l+1}^{z}),
\end{equation}
where $\sigma_l^x$, $\sigma_l^y$, $\sigma_l^z$ are three Pauli matrices for the $l$-th spin, and $J_l$ is the coupling strength between the $l$-th spin and the $(l+1)$-th spin.

The XXX Hamiltonian has the $SU(2)$ symmetry, whose Lie algebra is composed by the total spin component operators:
\begin{align}
  \label{eq:36}
  S_z & = \frac{1}{2} \sum_l \sigma^z_l, \\
  S_x & = \frac{1}{2} \sum_l \sigma^x_l, \\
  S_y & = \frac{1}{2} \sum_l \sigma^y_l.
\end{align}
Note that we take $\hbar=1$ throughout this paper. The bases  of the irreducible representations of the $SU(2)$ group can be specified by the common eigenstates of $S^2=S_x^2+S_y^2+S_z^2$ and $S_z$.

In addition, the XXX Hamiltonian has the complex conjugate symmetry, where the complex conjugate operator is denoted as $K$, which is an antiuniatry transformation. It is worthy to emphasize that we assume that the common eigenvectors of $\{\sigma^z_l, l\in\{1,2,\cdots,L\}\}$ are invariant under the transformation $K$.

Thus the symmetry group of the XXX Hamiltonian can be written as
\begin{equation}
G = \{R\} + \{K R\},
\end{equation} 
where $R\in SU(2)$. Here $\{R\}$ contains all unitary symmetry operations, which forms a subgroup of $G$, and $\{KR\}$ contains all antiunitary symmetry operations.

The XXX Hamiltonian has the $K$ symmetry, which implies that $\comm{K}{H_{XXX}}=0$. If $|\psi\rangle$ is an eigenstate of $H_{XXX}$ with energy $E$, then $H_{XXX}K |\psi\rangle= K H_{xxx}|\psi\rangle=  K E|\psi\rangle=EK|\psi\rangle$. Then $|\psi_R\rangle=\frac{1+K}{2} |\psi\rangle$ and $|\psi_I\rangle=\frac{1-K}{2i} |\psi\rangle$ also satisfy $H_{XXX}|\psi_R\rangle=E|\psi_R\rangle$, $H_{XXX}|\psi_I\rangle=E|\psi_I\rangle$. In addition, $K|\psi_R\rangle=|\psi_R\rangle$ and $K|\psi_I\rangle=|\psi_I\rangle$, i.e., $|\psi_R\rangle$ and $|\psi_I\rangle$ have real coefficients in the bases  of $\{\sigma^z_l\}$. Thus all the coefficients of the eigenvectors of $H_{XXX}$ in the above basis can be taken to be real.

In addition, the complex conjugate operation $K$ commutates with $S^2$ and $S_Z$, which implies that the common eigenvectors of $S^2$ and $S_z$ can always be taken as real in the $\{\sigma^z_l\}$ basis. This also implies that operation $K$ can not affect the dimension of irreducible representations of $G$, i.e., those of the symmetry group $G$. In these irreducible bases, $D_{nj}^p$ and $c_j^{p}$ in Eq.~\eqref{eq:12} are real, which implies that Eq.~\eqref{eq:16} does not give independent equations of $\{a_n\}$. Thus the number of equations are $\nu_p$, and Eq.~\eqref{eq:22} becomes
\begin{equation}
  \label{eq:37}
  \nu_p \ge N.
\end{equation}

Now we list the decompositions of the Hilbert space $\mathcal{H}$ into the irreducible subspaces $\mathcal{H}_p$ of $G$ for the XXX Hamiltonian with different $L$:
\begin{equation}
  \label{eq:38}
  \mathcal{H} = \oplus_p (\nu_p \times \mathcal{H}_p).
\end{equation}
which also implies that the equalities of the dimensions of corresponding Hilbert spaces: $d_{\mathcal{H}}=\sum_p \nu_p \times d_p$.

The specification of the bases of the irreducible representation for the symmetry group $G$
can be realized by giving the set of commutative Hermitian operators:
\begin{equation}
  \label{eq:2}
  \{S_z, S_n^2, n \in\{2,\cdots,L\} \}
\end{equation}
where $S_n$ is the total spin operator of the first $n$ spins, and $S_L=S$.

The list of such decompositions is given in TABLE~\ref{table:1}. For example, when $L=5$, the Hilbert space is the direct sum of $5$ $2$-dimensional Hilbert spaces, which can be decomposed into direct product of the following irreducible subspaces of the symmetry group $G$: $1$ $6$-dimensional irreducible subspace, $4$ $4$-dimensional irreducible subspaces, and $5$ $2$-dimensional irreducible subspaces.

\begin{table}[htbp]
  \begin{tblr}{cc}
  \toprule
	$L$ & Decomposition of Hilbert space\\
	\midrule
	2 & $\mathbf{2}^{\otimes 2}=\mathbf{3}\oplus\mathbf{1}$ \\
	3 & $\mathbf{2}^{\otimes 3}=\mathbf{4}\oplus(2\times\mathbf{2})$ \\
	4 & $\mathbf{2}^{\otimes 4}=\mathbf{5}\oplus(3\times\mathbf{3})\oplus(2\times\mathbf{1})$ \\
	5 & $\mathbf{2}^{\otimes 5}=\mathbf{6}\oplus(4\times\mathbf{4})\oplus(5\times\mathbf{2})$ \\
	6 & $\mathbf{2}^{\otimes 6}=\mathbf{7}\oplus(5\times\mathbf{5})\oplus(9\times\mathbf{3})\oplus(5\times\mathbf{1})$ \\
	7 & $\mathbf{2}^{\otimes 7}=\mathbf{8}\oplus(6\times\mathbf{6})\oplus(14\times\mathbf{4})\oplus(14\times\mathbf{2})$ \\
    \bottomrule
    \end{tblr}
  \caption{Distributions of irreducible subspaces of the symmetry group $G$ of the $L$-site XXX Hamiltonians for $L$ from $2$ to $7$.}
  \label{table:1}
\end{table}

In theory, we can decide whether the Hamiltonian can be recovered up to a constant from  a given eigen state $|\psi\rangle$ by the following steps: First, the site length $L$ is determined, the number of unknown parameters $\{a_n\}$ is given by $N=L-1$; Second, we decide which irreducible representation $|\psi\rangle$ belongs to, which can be derived by calculating $\langle\psi|S^2|\psi\rangle$ to get $S$, then $d_p=2S+1$, looking for $\nu_p$ in the table; Third, we gives the conclusion by the rule: if $\nu_p\ge N$, then the Hamiltonian learning successes; otherwise it fails.

According to the above approach, we present the results on the Hamiltonian recovery of the XXX Hamiltonian model in TABLE~\ref{tab:1a}, where $O$ denotes the case that the Hamiltonian recovery successes, and $X$ denotes we can not recover the original Hamiltonian from a single eigen energy state.

\begin{table}[htbp]
  \centering
 \begin{tblr}{c|cccccccc}
  \toprule
  \diagbox{$L$}{$S$} & $0$ & $\frac{1}{2}$ & $1$ & $\frac{3}{2}$ & $2$ & $\frac{5}{2}$ & $3$ & $\frac{7}{2}$ \\
  \midrule
  3 & / & O & / & X & / & / & / & / \\
  4 & X & / & O & / & X & / & / & / \\
  5 & / & O & / & O & / & X & / & / \\
  6 & O & / & O & / & O & / & X & / \\
  7 & / & O & / & O & / & O & / & X \\
  \bottomrule
\end{tblr}
  \caption{Recovery of the XXX Hamiltonian of a spin chain with different length $L$ from an eigen state $|\psi\rangle$ belonging to different $S$ irreducible representations.}
  \label{tab:1a}
\end{table}

We also verifies the validness of the above approach by numerical algorithms discussed in Eq.~\eqref{eq:32}.

\subsection{The XXZ Spin Chain}\label{sec:xxz-spin-chain}

The second example we consider is a $1$-dimensional spin-$\frac{1}{2}$ chain with the XXZ Hamiltonian given by
\begin{equation}
    H_{XXZ} = \sum_{l=1}^{L-1}J_l^z\sigma_{l}^{z}\sigma_{l+1}^{z}+J_l^{xy}(\sigma_{l}^{x}\sigma_{l+1}^{x}+\sigma_{l}^{y}\sigma_{l+1}^{y}),
  \end{equation}
  where $J_l^z$ and $J_l^{xy}$ are unknown real parameters, and the number of the unknown parameters $N=2L$.

The system described by the XXZ Hamiltonian has the $U(1)$ symmetry with the generator $S_z$ defined by Eq.~\eqref{eq:36}. In addition, the system has two discrete symmetries: One unitary symmetry operator $\Pi_x=\otimes_l \sigma^x_l$ and the complex conjugate symmetry operator $K$.    The symmetry group $G$ of the XXZ Hamiltonian can be written as
\begin{equation}
G = \{U\}+\{K U\},
\end{equation}
where $\{U\}$ a unitary group generated by the $U(1)$ group and unitary operator $\Pi_x$.

Note that $\Pi_x^2=I$, and $\Pi_x$ is also a Hermitian operator with eigenvalues $\pm 1$. In addition, note that
\begin{equation}
  \label{eq:39}
  \Pi_x S_z \Pi_x = - S_z,
\end{equation}
which implies
\begin{equation}
  \label{eq:40}
  S_z \Pi_x |i,m_z\rangle = - m_z \Pi_x |i, m_z\rangle
\end{equation}
where $S_z|i,m_z\rangle=m_z|i,m_z\rangle$ and the index $i$ denotes the degeneracy. If $m_z\neq 0$, then $|i,m_z\rangle$ and $\Pi |i,m_z\rangle$, which spans a $2$-dimensional irreducible subspace for the unitary subgroup $\{U\}$. If $m_z=0$, then $|i,0\rangle$ and $\Pi_x|i,0\rangle$ are degenerate for $S_z$, we can construct $|i,0,\pm1\rangle=\frac{1\pm\Pi_x}{2}|i,0\rangle$, which are the commen eigenstates for $S_z$ and $\Pi_x$ with eigen values $0$ and $\pm 1$ respectively. Obviously, the two states $|i,0,\pm 1\rangle$ are $1$-dimensional irreducible subspace for the unitary subgroup $\{U\}$. Let us denote the $2$-dimensional irreducible representation with $m_z\neq 0$ as $2^{\abs{m_z}}$, and denote the $1$-dimensional irreducible representation as $1^{0, +1}$ or $1^{0, -1}$.

Note that the $XXZ$ Hamiltonian $H_{XXZ}$, the $z$-component of the total spin $S_z$, and the symmetry operator $\Pi_x$ are invariant under the action of the complex conjugate transformation $K$. With the similar argument in the $XXX$ model, we will also obtain Eq.~\eqref{eq:37}. In addition, the irreducible subspaces of the symmetry group $G$ are the same as those of $\{U\}$.

The specification of the symmetry group $G$ for the XXZ Hamiltonians can be given by the following set of commutative Hermitian operators:
\begin{equation}
  \label{eq:3}
  \{\abs{S_z}, \Pi_x, \sigma_1^z \sigma_i^z, \; i\in\{2,3,\cdots,L\} \}
\end{equation}
Note that their common eigenstates gives a basis of all the irreducible representations of $G$.

Now we are ready to present TABLE.~ \ref{table:5} to show the decomposition of the Hilbert space according the irreducible representations of the symmetry group $G$ of the XXZ Hamiltonian, where the degeneracy of any irreducible representation is also given in the form of Eq.~\eqref{eq:38}.

\begin{widetext}
  \begin{center}
  \begin{table}[h]
    \begin{tblr}{cc}

      \toprule
	$L$ & Decomposition of Hilbert space\\
	\midrule
	2 & $\mathbf{2}^{\otimes 2}=\mathbf{1}^{0,1}\oplus\mathbf{1}^{0,-1}\oplus\mathbf{2}^{ 1}$ \\
	3 & $\mathbf{2}^{\otimes 3}=\mathbf{2}^{\frac{3}{2}}\oplus(3\times\mathbf{2}^{\frac{1}{2}})$ \\
	4 & $\mathbf{2}^{\otimes 4}=(3\times\mathbf{1}^{0,1})\oplus(3\times\mathbf{1}^{0,-1})\oplus(4\times\mathbf{2}^{ 1})\oplus\mathbf{2}^{ 2}$ \\
	5 & $\mathbf{2}^{\otimes 5}=\mathbf{2}^{\frac{5}{2}}\oplus(5\times\mathbf{2}^{\frac{3}{2}})\oplus(10\times\mathbf{2}^{\frac{1}{2}})$ \\
	6 & $\mathbf{2}^{\otimes 6}=(10\times\mathbf{1}^{0,1})\oplus(10\times\mathbf{1}^{0,-1})\oplus(15\times\mathbf{2}^{ 1})\oplus 6\times\mathbf{2}^{ 2}\oplus\mathbf{2}^{ 3}$ \\
	7 & $\mathbf{2}^{\otimes 7}=\mathbf{2}^{\frac{7}{2}}\oplus(7\times\mathbf{2}^{\frac{5}{2}})\oplus(21\times\mathbf{2}^{\frac{3}{2}})\oplus(35\times\mathbf{2}^{\frac{1}{2}})$ \\
      \bottomrule
    \end{tblr}
 \caption{\enspace Decomposition of the Hilbert space into the irreducible subspaces of the symmetry group $G$ of the $L$-site XXZ Heisenberg model.}
  \label{table:5}
\end{table}
  \end{center}

\end{widetext}

Simulations of Hamiltoninan learning are performed on $H_{XXZ}$, all the cases for the Hamiltonian recovery for chain lengths ranging from 2 to 7 are presented in Table~\ref{table:6}.

\begin{table}[htbp]
  \begin{tblr}{c|cccccccc}
    \toprule
	\diagbox{$L$}{$\abs{S^z}$} & 0&$\frac{1}{2}$&$ 1$&$\frac{3}{2}$&$ 2$&$\frac{5}{2}$&$ 3$&$ \frac{7}{2}$\\
	\midrule
	2 & X&/&X&/&/&/&/&/ \\
        3 & /&X&/&X&/&/&/&/\\
        4 & X&/&X&/&X&/&/&/\\
        5 & /& O &/&X&/&X&/&/\\
        6 &  O&/& O&/&X&/&X&/\\
        7 & /& O&/& O &/&X&/&X\\
      \bottomrule
    \end{tblr}
  \caption{\enspace Recovery of XXZ Hamiltonian from its eigenstate for $L$ from $2$ to $7$, where $O$ denotes the recovery ``succeed'', and ``X'' denotes the recovery ``fail''.}
  \label{table:6}
\end{table}

We observe that the results on the Hamiltonian recovery in Table~\ref{table:6} are consistent with the Hamiltonian recovery criteria~\eqref{eq:37}.

\subsection{Accidental degeneracy: From the XXZ Hamiltonian to the XXX Hamiltonian}
\label{sec:accid-degen-from}

When we study the recovery of the XXZ Hamiltonian, accidentally we will meet an XXX Hamiltonian, which is the case if $J^{xy}_l=J^z_l$ for any $l$. In this case, the eigen state must be the one we discussed in Sec.~\ref{sec:xxx-spin-chain}, but not the eigen state in Sec.~\ref{sec:xxz-spin-chain}. In this case, the neccessary condition of recovering the Hamiltonian up to a constant becomes
\begin{equation}
  \label{eq:41}
  \sum_{p\in \Lambda(\psi)} \nu_p \ge N,
\end{equation}
where the degeneracy $\nu_p$ are list in TABLE~\ref{table:5}, and $N=2(L-1)$. Here the index $p$ is determined by $\abs{m}$, and $\nu_p$ also depends on the chain length $L$. The fact that $\Lambda(\psi)$ contains more than one irreducible representations of the XXZ Hamiltonian is a signature of the appearance of accidental degeneracy. In this case, we can not make a precise prediction of the Hamiltonian recoverability based on incomplete information of symmetry.

Now the energy eigen state $|\psi\rangle$ belongs to the irreducible representation labeled by the total spin $S$, whose irreducible subspace dimension is $d_S=2S+1$. For a given $S$, $m$ takes values in the set $\{-S, -S+1, \cdots, S\}$. Thus the recovery condition~\eqref{eq:41} becomes
\begin{equation}
  \label{eq:42}
  \sum_{\abs{m}\le S}\nu_{\abs{m}}(L) \ge 2(L-1).
\end{equation}
According to Eq.~\eqref{eq:42}, we present the prediction results on the recovery of the XXX Hamiltonian as an accidental case of the recovery of the XXZ Hamiltonian in TABLE~\ref{tab:2a}. It is worthy to point out that Eq.~\eqref{eq:42} is only a necessary condition for recoverability.

\begin{table}[htbp]
  \centering
 \begin{tblr}{c|cccccccc}
  \toprule
  \diagbox{$L$}{$S$} & $0$ & $\frac{1}{2}$ & $1$ & $\frac{3}{2}$ & $2$ & $\frac{5}{2}$ & $3$ & $\frac{7}{2}$ \\
  \midrule
  2 & X & / & O & / & / & / & / & / \\
  3 & / & X & / & O & / & / & / & / \\
  4 & X & / & O & / & O & / & / & / \\
  5 & / & O & / & O & / & O & / & / \\
  6 & O & / & O & / & O & / & O & / \\
  7 & / & O & / & O & / & O & / & O \\
  \bottomrule
\end{tblr}
  \caption{The necessary condition on the recovery of the XXX Hamiltonian of a spin chain with different length $L$ from an eigen state $|\psi\rangle$ belonging to different $S$ irreducible representations as an accidental case of revovery of the XXZ Hamiltonian.}
  \label{tab:2a}
\end{table}

The numerical results on the recovery of the XXX Hamiltonian as an accidental case of the recovery of the XXZ Hamiltonian in TABLE~\ref{tab:2c} and TABLE~\ref{tab:2b}. The results on the recovery in TABLE~\ref{tab:2b} are directly conclusion from the possible ranks of matrices $Q$'s in TABLE~\ref{tab:2c}.

 \begin{table}[htbp]
  \centering
 \begin{tblr}{c|cccccccc}
  \toprule
  \diagbox{$L$}{$S$} & $0$ & $\frac{1}{2}$ & $1$ & $\frac{3}{2}$ & $2$ & $\frac{5}{2}$ & $3$ & $\frac{7}{2}$ \\
  \midrule
	2 & 1&/&1&/&/&/&/&/ \\
        3 & /&3&/&1,3&/&/&/&/\\
        4 & 3&/&3,4,6&/&1,3,4&/&/&/\\
        5 & /&8&/&5,8&/&1,5&/&/\\
        6 &  10&/& 10&/&6,10&/&1,5,6&/\\
        7 & /& 12&/&12&/&7,12&/&1,7\\
  \bottomrule
\end{tblr}
  \caption{The possible rank of constrained matrix $Q$ on the recovery of the XXX Hamiltonian of a spin chain with different length $L$ from an eigen state belonging to different $S$ irreducible representations as an accidental case of recovery of the XXZ Hamiltonian.}
  \label{tab:2c}
\end{table}

\begin{table}[htbp]
  \centering
 \begin{tblr}{c|cccccccc}
  \toprule
  \diagbox{$L$}{$S$} & $0$ & $\frac{1}{2}$ & $1$ & $\frac{3}{2}$ & $2$ & $\frac{5}{2}$ & $3$ & $\frac{7}{2}$ \\
  \midrule
  2 & X & / & X & / & / & / & / & / \\
  3 & / & X & / & X & / & / & / & / \\
  4 & X & / & OX & / & X & / & / & / \\
  5 & / & O & / & OX & / & X & / & / \\
  6 & O & / & O & / & OX & / & X & / \\
  7 & / & O & / & O & / & OX & / & X \\
  \bottomrule
\end{tblr}
  \caption{The numerical results on the recovery of the XXX Hamiltonian of a spin chain with different length $L$ from an eigen state $|\psi\rangle$ belonging to different $S$ irreducible representations as an accidental case of revovery of the XXZ Hamiltonian.}
  \label{tab:2b}
\end{table}

Comparing the results of TABLE~\ref{tab:2a} with those of TABLE~\ref{tab:1a}, we find that all the recoverable cases in TABLE~\ref{tab:1a} are still recoverable in TABLE~\ref{tab:2a}. Logically, the best results we can obtain are those in TABLE~\ref{tab:1a} with the whole knowledge of the symmetry. In fact, the numerical results in TABLE~\ref{tab:2b} are close to those in TABLE~\ref{tab:1a}, except that some recoverable cases in TABLE~\ref{tab:1a} become partially recoverable, which are denoted by $OX$.

\section{conclusions}

It is worthy to point out that without symmetry analysis of the family of Hamiltonians, we still can apply the previous methods to solve the recoverabilty of Hamiltonian with any given eigenstate by the previous methods\cite{PhysRevX.8.031029,PhysRevLett.122.020504,Hou_2020,PhysRevA.105.012615}. The power of the symmetry analysis lies in how to find out all the really independent linear equations for the unkonwn parameters in the symmetric Hamiltonian to be recovered, and also gives a practical criterion on the recoverability of Hamiltonian without accidental symmetry via the symmetry of one of its steady state.

In summary, we investigate the problem of when a local Hamiltonian can be uniquely recovered up to a constant from its eigenstate, which extends the previous work to the realm of symmetric Hamiltonians. Our work introduces the symmetry group of the Hamiltonian family with unknown parameters. By applications of group representation theory,  we shows that without accidental symmetry, the largest number of linearly independent equations derived from an eigenstate equals to the degeneracy of irreducible representation of Hamiltonian symmetry group that the state belongs to. In the cases when accidental symmetry appears, we find the eigen state may belongs to different irreducible representations of the known symmetry group, which makes precise prediction of the Hamiltonian recovery impossible due to the lack of complete information on symmetry. We also demonstrate our general approach with two $1$-dimensional spin chain models: the XXX Hamiltonian model and the XXZ Hamiltonian model. The theoretical results on the Hamiltonian recovery are verified with numerical simulations.

An open question is how to extend our results to more complex Hamiltonians, such as those that involve all‑to‑all two‑body interactions or describe indistinguishable particles. Moreover, we would like to know whether our findings remain valid for open quantum systems, and what limitations exist on the measurements required to extract steady‑state information. Addressing these issues is crucial for the practical implementation of Hamiltonian learning in the pursuit of novel quantum materials.

\begin{acknowledgements}
This work is supported by Science Challenge Project (Grant No. TZ2025017), National Key Research and Development Program of China (Grant No. 2021YFA0718302 and No. 2021YFA1402104), National Natural Science Foundation of China (Grants No. 12075310).
\end{acknowledgements}

\bibliographystyle{apsrev4-2} \bibliography{tomo}

\end{document}